\documentclass[zamm,fleqn,a4paper]{w-art}
\usepackage{times}
\usepackage{w-thm}
\theoremstyle{plain}

\theoremstyle{definition}

\usepackage[]{epsfig,graphicx,amssymb}
\begin{document}
\DOIsuffix{theDOIsuffix}
\Volume{XX}
\Issue{1}
\Month{01}
\Year{2003}
\pagespan{1}{}
\Receiveddate{\today}
\Reviseddate{~}
\Accepteddate{~}
\Dateposted{~}
\keywords{Duffing oscillator, Melnikov criterion, chaotic
vibration}
\title[Vibration of Generalized Double Well Oscillators]{
Vibration of Generalized Double Well Oscillators
}

\author[G. Litak]{Grzegorz Litak\inst{1,2}%
  \footnote{Corresponding
author,~e-mail:~\textsf{g.litak@pollub.pl},
            Phone: +4881\,538\,1573,
            Fax: +4881\,525\,0808}}

\author[M. Borowiec]{Marek Borowiec\inst{1,}\footnote{m.borowiec@pollub.pl}}

\address[\inst{1}]{Department of Applied Mechanics, Technical University of
Lublin, Nadbystrzycka 36, PL-20-618 Lublin, Poland}

\address[\inst{2}]{Institut f\"{u}r Mechanik und Mechatronik, Technische
Universit\"{a}t Wien,
 Wiedner
Hauptstra$\beta$e
8 - 10 A-1040 Wien, Austria}

\author[A. Syta]{Arkadiusz Syta\inst{3,}\footnote{a.syta@pollub.pl}}

\address[\inst{2}]{Department of Applied
Mathematics, Technical University of Lublin, Nadbystrzycka 36, PL-20-618 Lublin, Poland}

\begin{abstract}
 We have applied the Melnikov criterion to examine a global homoclinic bifurcation and 
transition 
to chaos 
 in a case of a double well dynamical
system with a nonlinear fractional damping term and external excitation. 
The usual double well Duffing potential having 
a negative square term and positive quartic term
has been generalized to a double well potential with 
a negative square term and a positive one 
with an arbitrary real exponent $q > 2$. 
We have also used a fractional damping term with an arbitrary power $p$ applied to  velocity 
which
enables one to cover a wide range of realistic damping factors: from 
dry friction $p \rightarrow 0$ to turbulent resistance phenomena $p=2$.
  Using perturbation methods we have found a critical forcing amplitude $\mu_c$ 
above which the 
system may 
 behave chaotically.
Our results show that the vibrating system is less stable in  transition to chaos
for smaller $p$ satisfying
an exponential scaling low. The critical amplitude $\mu_c$ as an exponential 
function of $p$.  
 The analytical results have  been 
illustrated by numerical simulations using standard nonlinear tools such as 
 Poincare maps and the maximal Lyapunov 
 exponent. As usual for chosen system parameters 
we have identified a chaotic motion
above the critical Melnikov amplitude $\mu_c$.  
\end{abstract}
\maketitle

\section{Introduction}
A nonlinear oscillator with single or double well potentials 
of the Duffing type and linear damping is   one of 
the simplest 
systems leading to chaotic 
motion studied by \cite{Ueda1979,Ueda1980,Ueda1981,Moon1979}.
The problem of  its nonlinear vibrations   
has attracted researchers from various 
 fields of research across natural science and physics  
\cite{Moon1979,Zalalutdinov2003,Chong2004},  mathematics 
\cite{Guckenheimer1983}  mechanical 
engineering 
\cite{Szemplinska1993,Szemplinska1995,Moon1987,Tyrkiel2005,Litak2006a}; 
and finally
electrical 
engineering \cite{Ueda1979,Ueda1980,Ueda1981}.
This system, for a negative linear part of stiffness, shows homoclinic orbits, 
and the transition to chaotic vibration can be treated 
analytically
via the Melnikov method \cite{Melnikov1963}. 
Such a treatment has been already
 performed successfully to selected problems with various potentials  
\cite{Guckenheimer1983,Wiggins1990,Moon1987}.
Vibrations of a single Duffing oscillator have got a large bibliography
\cite{Ueda1979,Ueda1980,Ueda1981,Moon1979,Zalalutdinov2003,Chong2004,Guckenheimer1983,Wiggins1990,Szemplinska1993,Szemplinska1995,Tyrkiel2005,Litak2006a,Moon1987,Litak1999,Trueba2000,Trueba2002,Borowiec2007}.
In the last decade 
coupled Duffing oscillators  
\cite{Warminski1999,Warminski2000,Lifshitz2003,Maccari2002,Maccari2003}
with numerous 
modifications to potential and
forcing parts have been studied.
On the other hand the problem of nonlinear
 damping in chaotically vibrating system has not been discussed in detail. Some 
insight into this problem can also be found   
in the context of self excitation effects
 
\cite{Litak1999,Warminski1999,Maccari2002,Maccari2003,Awrejcewicz1986,Litak2006b,Siewe2004}.
and dry friction effects 
\cite{Brockley1970,Ibrahim1994,Galvanetto1999,Leine2000}
In the paper by Trueba {\em et al.} \cite{Trueba2000}, 
the systematic discussion
 on square and cubic damping effects on global homoclinic bifurcations in the 
Duffing system has been given. Recently Trueba {\em et al.} \cite{Trueba2002} and 
Borowiec {\em et al.} 
\cite{Borowiec2007} have analyzed
a single degree of freedom nonlinear oscillator with the Duffing
potential and fractional damping.
Different aspects  fractionally damped systems have been studied recently by 
Mickens {\em et al.}, Gottlieb, and Mickens \cite{Mickens2003a,Gottlieb2003,Mickens2003b}. 
On the other hand Maia {\em et al.} and 
Padovan and Sawicki
\cite{Maia1998,Padovan1998,Sheu2007} analyzed similar systems  where fractional 
damping have 
been 
introduced 
in different way through a fractional derivative.
Awrejcewicz and Holicke \cite{Awrejcewicz1999} and more recently 
Awrejcewicz and Pyryev \cite{Awrejcewicz2006} applied Melnikov's method in 
the presence of dry friction
for a stick-slip oscillator.
More general introduction to the  problem of non-smooth or discontinuous mechanical systems 
can be found in \cite{Awrejcewicz2003,Kunze2001}.

In the present paper we revisit this problem 
looking for a global 
homoclinic bifurcation and transition to chaotic vibrations in a system described 
by a more general double well potential where its usual positive quartic term
has been generalized to term 
 with an arbitrary real exponent grater than 2. Below we would also apply a nonlinear 
damping  term with a fractional exponent
covering the gap between viscous, dry friction and turbulent damping phenomena. 

\begin{figure}[htb]

\centerline{
\epsfig{file=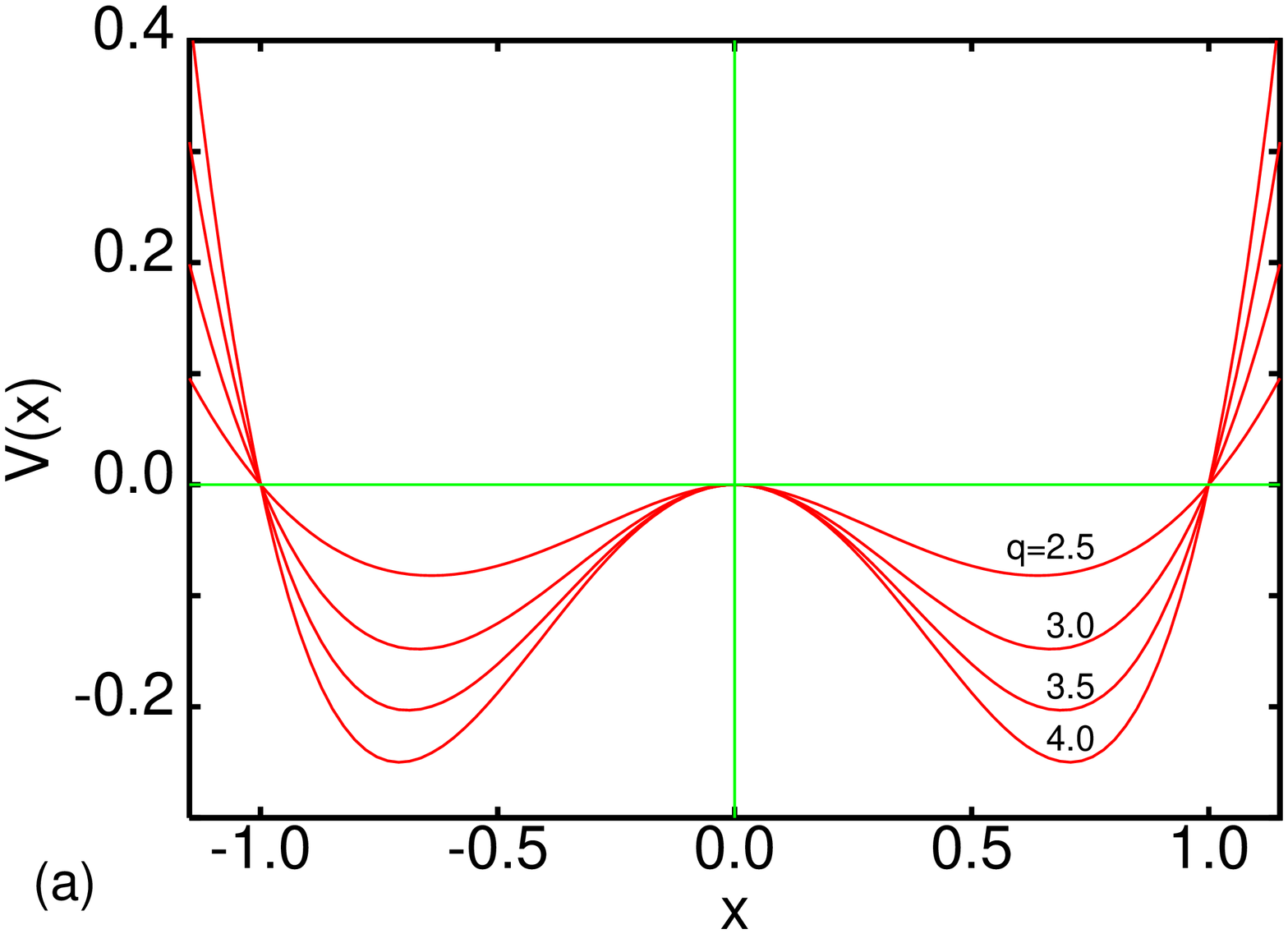,width=5.5cm,angle=-90} \hspace{-0.3cm}
\epsfig{file=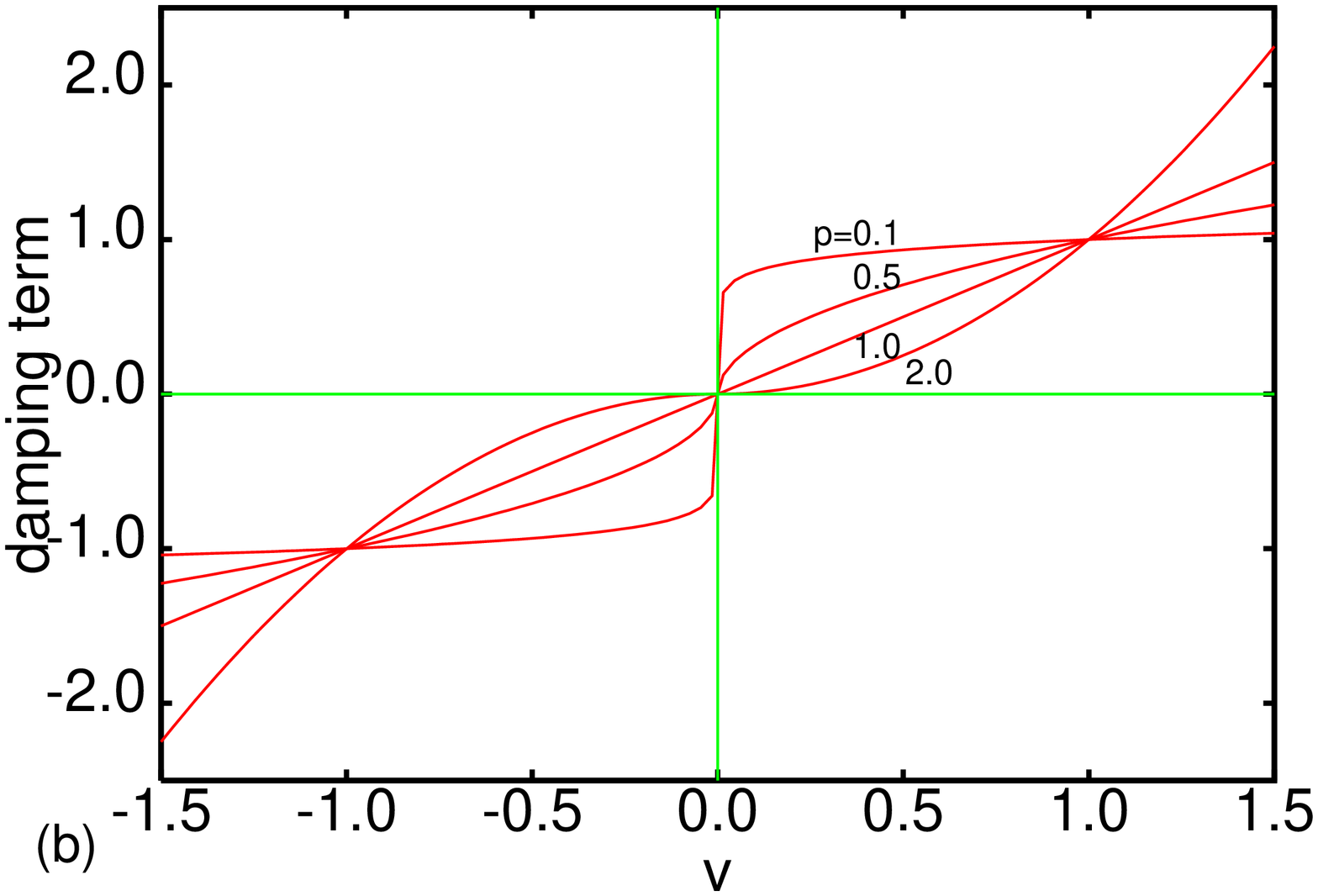,width=5.5cm,angle=-90}}

\caption{ \label{fig1}
External potential $V(x)=\frac{\delta x^2}{2}+\frac{\gamma|x|^q}{q}$ (Eq. 3) for $\delta=-2$
for a few values of $q$ ($q=\gamma > 2$ in Fig. 1a),
Damping term for various $p$ (Fig. 1b).
}
 \end{figure}

The equation of motion has the following form:
\begin{equation}
\label{eq1}
\ddot{x} +\alpha \dot{x} \left| \dot{x} \right|^{p-1}
+ \delta x +\gamma  {\rm sgn}(x)|x|^{q-1}=\mu \cos{ \omega t},
\end{equation}
where $x$ is displacement and $\dot{x}$ velocity, respectively, while
the external  force $F_x$: 

\begin{equation}
\label{eq2}
F_x= -\delta x -\gamma {\rm sgn}(x)|x|^{q-1},
\end{equation}
and corresponding potential $V(x)$ (Fig. \ref{fig1}a) is defined as:
\begin{equation}
\label{eq3}
V(x)=\frac{\delta x^2}{2} + 
\frac{\gamma |x|^q}{q},
\end{equation}
where $q > 2$ is a real number. In spite of the definition $V(x)$ (Eq. \ref{eq3}) in 
terms of absolute 
value $|x|$
it is still a function of $C^2$ class if only $q > 2$ (see Appendix A).

The non-linear damping term is defined by the exponent $p$:
\begin{equation} 
\label{eq4}
{\rm dpt}(\dot{x})=\alpha \dot{x} \left| \dot{x} \right|^{p-1}.
\end{equation}
In Fig. \ref{fig1}b
we have plotted the above function versus velocity ($v=\dot{x}$)  for few values of $p$.
Note that,
the case
$p \rightarrow 0$ (see $p=0.1$ in Fig. \ref{fig1}b  for a relatively small
velocity) mimics the
dry
friction phenomenon \cite{Brockley1970,Ibrahim1994}.

\section{Melnikov Analysis}

We start our analysis with the
unperturbed Hamiltonian:
$H^0$
\begin{equation}
\label{eq5}
H^0= \frac{v^2}{2} + V(x).
\end{equation}
Note that for our choice of potentials $\delta=-2$ and $\gamma=q$
(Fig. \ref{fig1}a)
$V(x)$  has the three nodal points ($x=-1$, 0, 1) where the middle one ($x=0$)
corresponds to the local peak at
the saddle point.
The existence of this point with a horizontal tangent enables occurrence of
homoclinic bifurcations. This includes transitions from regular to
chaotic solutions.  To study the effects of damping and excitation on the saddle point 
bifurcations, we apply small
perturbations around the
homoclinic orbits.
Our strategy is to use a small parameter $\epsilon$ to
the Eq. \ref{eq1} with perturbation terms.
Uncoupling Eq. \ref{eq1} into two differential equations of the first
order we obtain
\begin{eqnarray}
\label{eq6}
\dot{x} &=& v \\
\dot{v} &=& 
-\epsilon \tilde{\alpha} v \left| v \right|^{p-1}
- \delta x -\gamma {\rm sgn}(x) |x|^{q-1}+\epsilon \tilde{\mu} \cos{ \omega t}, \nonumber
\end{eqnarray}
where $\epsilon \tilde{\alpha}=\alpha$ and $\epsilon \tilde{\mu}=\mu$, respectively.

\begin{figure}[htb]
\centerline{
\epsfig{file=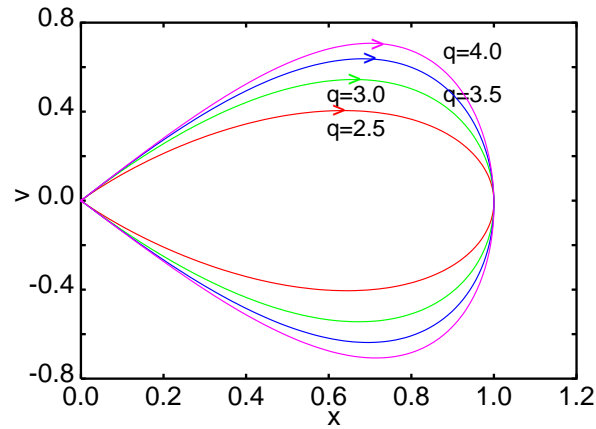,width=6.5cm,angle=-90}}

\caption{ \label{fig2}
Left hand side homoclinic orbits for unperturbed Hamiltonian (Eq. 
\ref{eq5}). 
Note, in our 
case the potential has reflection symmetry over 0-$y$ axis so the orbits    
appear in pairs for corresponding regions $x> 0$ and $x < 0$.}
 \end{figure}

At the saddle point $x=0$, for an unperturbed system (Fig. \ref{fig1}a), the system
velocity reaches
zero $v=0$
(for infinite time $t=\pm \infty$) so the
total energy has only its
potential part which has been gauged out to zero too.
Thus transforming Eqs. \ref{eq3} and \ref{eq5} for a nodal energy ($E=0$)
and for $\delta < 0$, $\gamma > 0$ we get the
following expression for velocity:
\begin{equation}
\label{eq7}
v= \frac{{\rm d} x}{{\rm d} t} =
\sqrt{2 \left(-
\frac{\delta x^2}{2} - \frac{\gamma
|x|^q}{q}\right)}.
\end{equation}

Performing integration  over $x$ we get
\begin{equation}
\label{eq8}
t-t_0= \pm \int  \frac{ {\rm d} x}{x\sqrt{
-\delta  - \frac{ 2 \gamma
|x|^{q-2}}{q}}},
\end{equation}
where $t_0$ represents here a time
like integration constant.

Integration in Eq. \ref{eq8} has been  performed analytically.
For $q > 2$, one can  write $x^*$ as:
\begin{equation}
\label{eq9}
x^*= x^*(t-t_0) =\pm \left(\frac{- \delta q}{2 \gamma}
\right)^{\frac{1}{q-2}}
\frac{1}{\cosh^{\frac{2}{q-2}}
\left[ \frac{(q-2)}{2}
\sqrt{-\delta} (t-t_0)
\right]}.
\end{equation}

The corresponding velocity $v^*$ reads:
\begin{equation}
\label{eq10}
v^*= v^*(t-t_0) =\mp
\sqrt{-\delta}\left(\frac{- \delta q}{2 \gamma}
\right)^{\frac{1}{q-2}}
\frac{\tanh \left[  \frac{(q-2)}{2}
\sqrt{-\delta} (t-t_0)
\right]}{\cosh^{\frac{2}{q-2}}
\left[ \frac{(q-2)}{2}
\sqrt{-\delta} (t-t_0)
\right]}.
\end{equation}
  
\begin{figure}[htb]
\centerline{
\epsfig{file=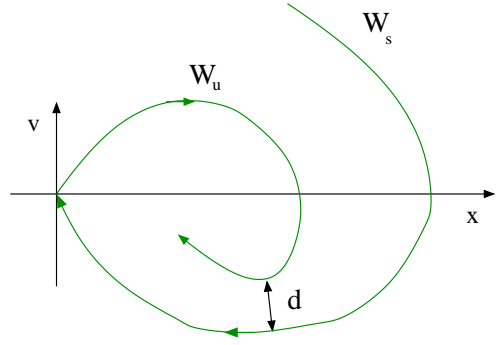,width=6.5cm,angle=0}}

\caption{ \label{fig3} Schematic plot of stable and unstable manifolds ($W_s$ and
$W_u$) of perturbed system Eq. \ref{eq6}.
$d$ denotes the distance between manifolds given by Melnikov function
$M(t_0)$ Eq. \ref{eq11}.}

 \end{figure}

\begin{figure}[htb]
\vspace{-0.7cm}
\centerline{
\epsfig{file=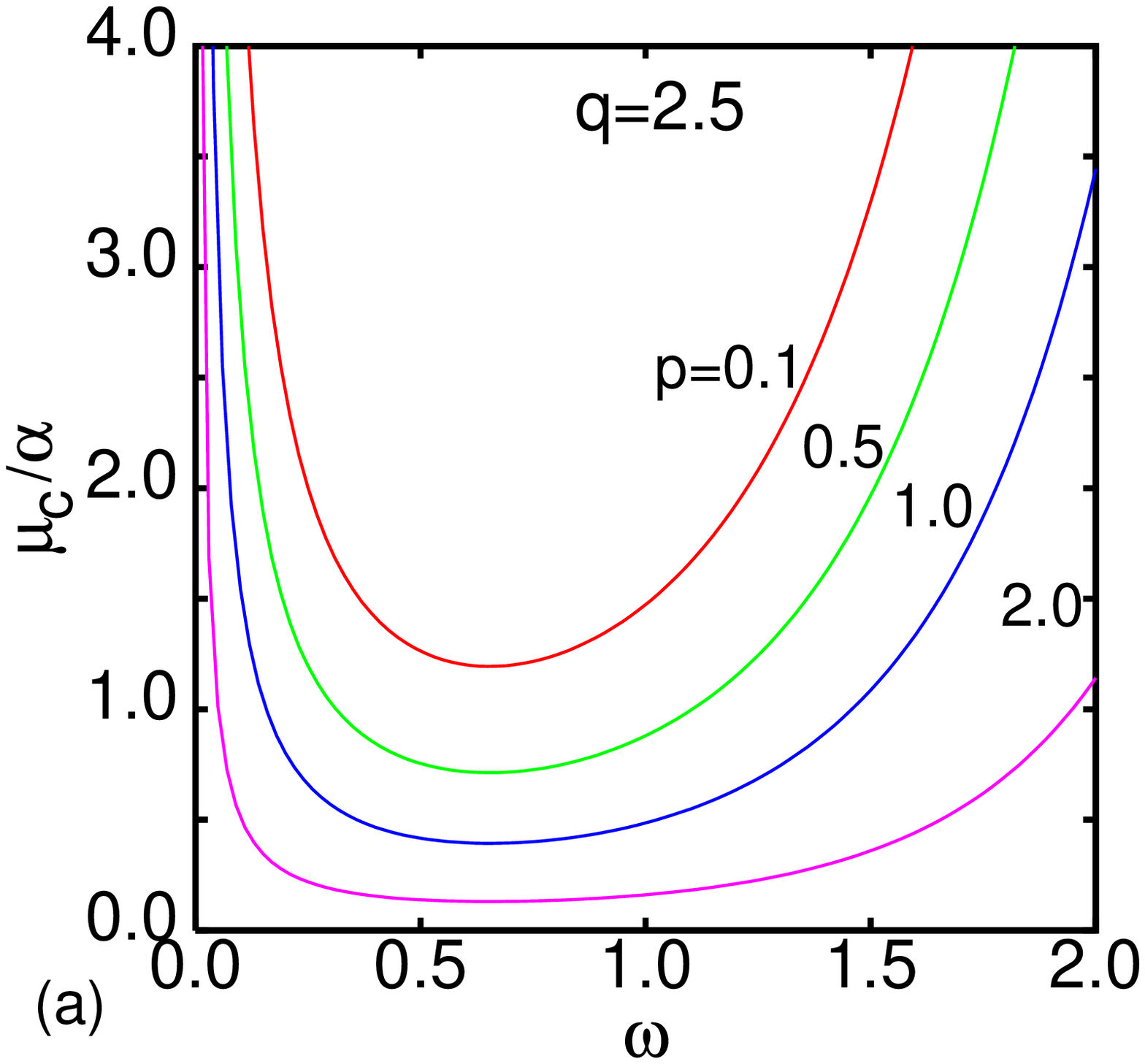,width=6.2cm,angle=-90} \hspace{0.3cm} 
\epsfig{file=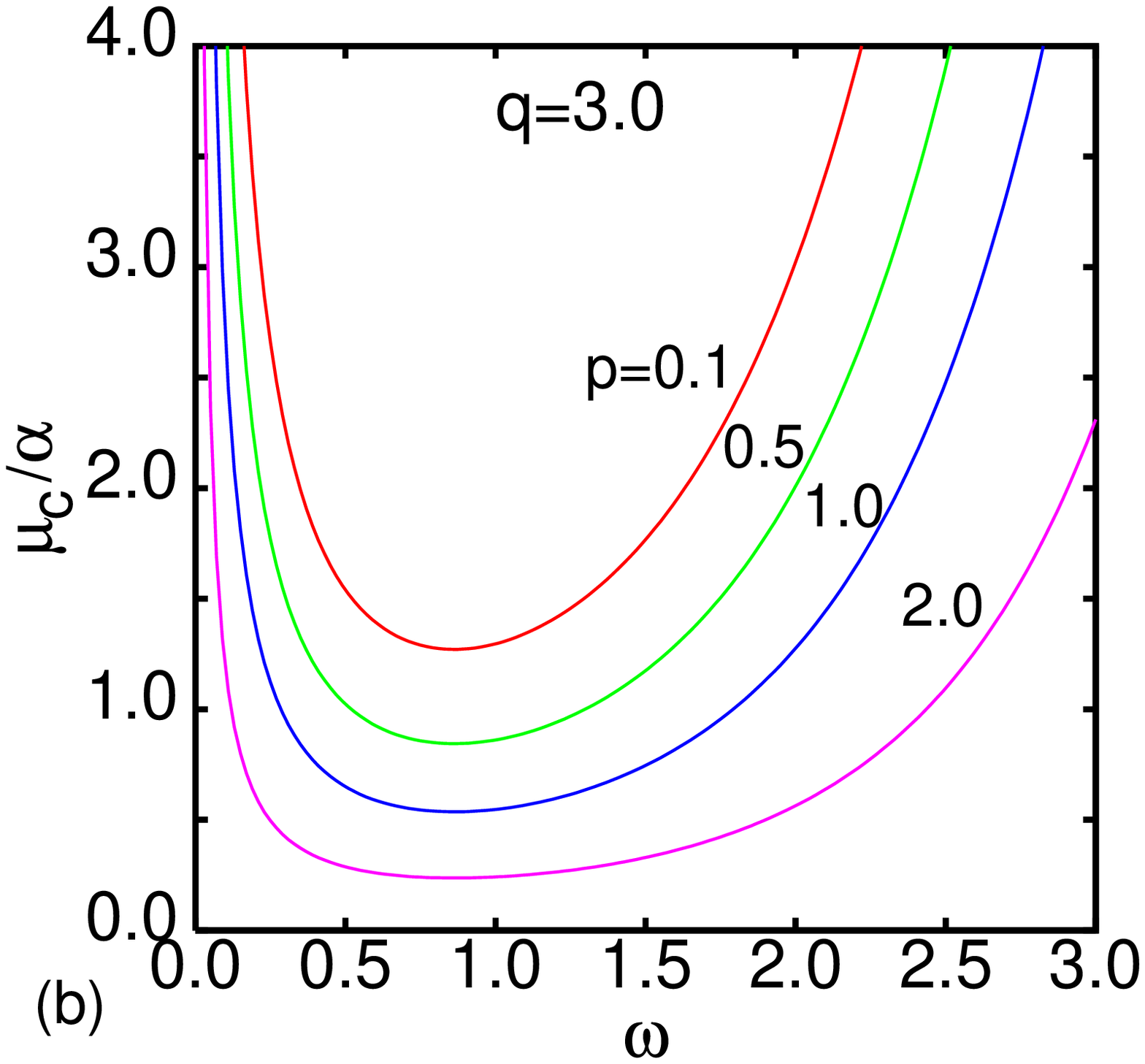,width=6.2cm,angle=-90}}

\vspace{-0.7cm}
\centerline{ 
\epsfig{file=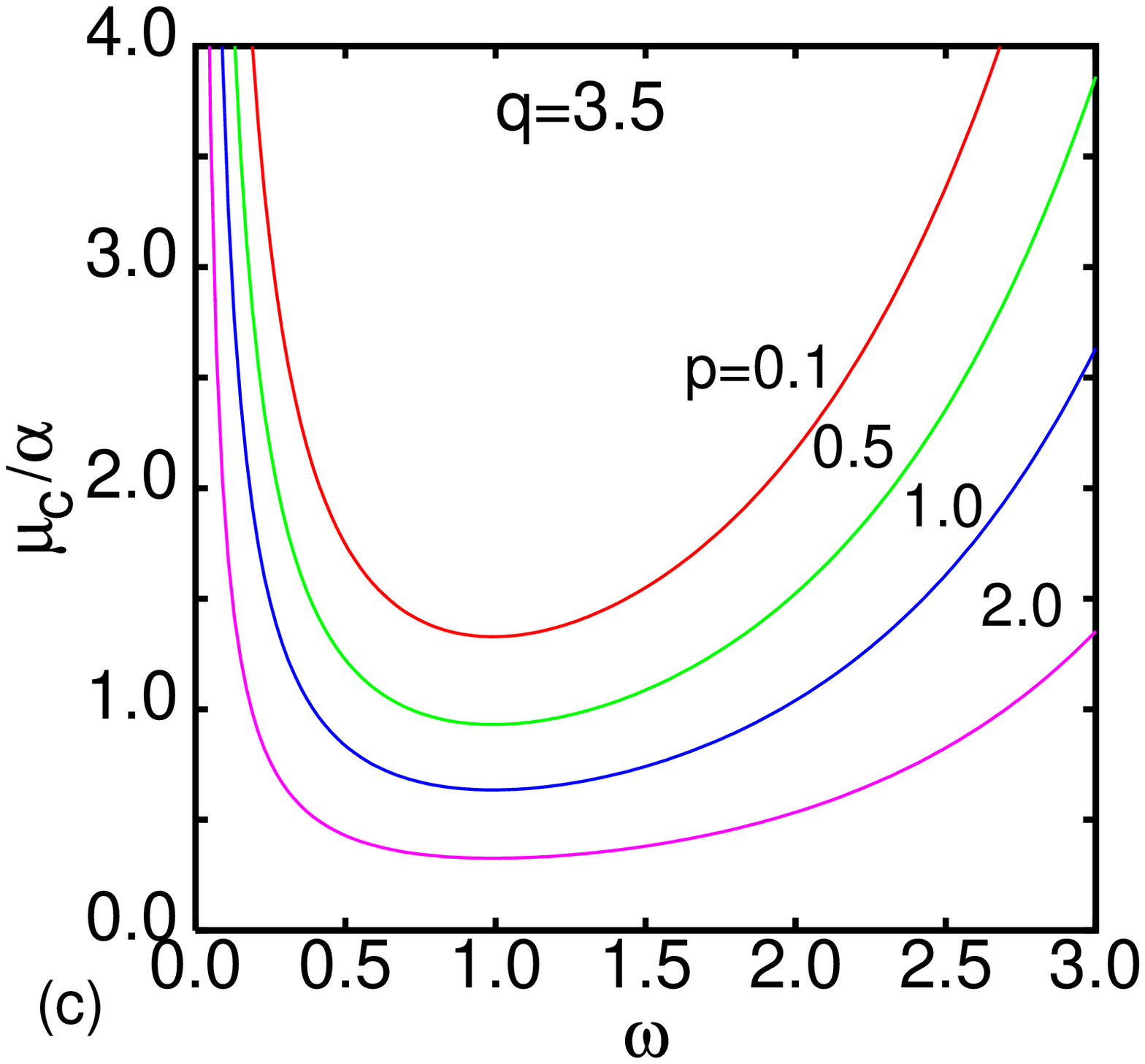,width=6.2cm,angle=-90} \hspace{0.3cm}
\epsfig{file=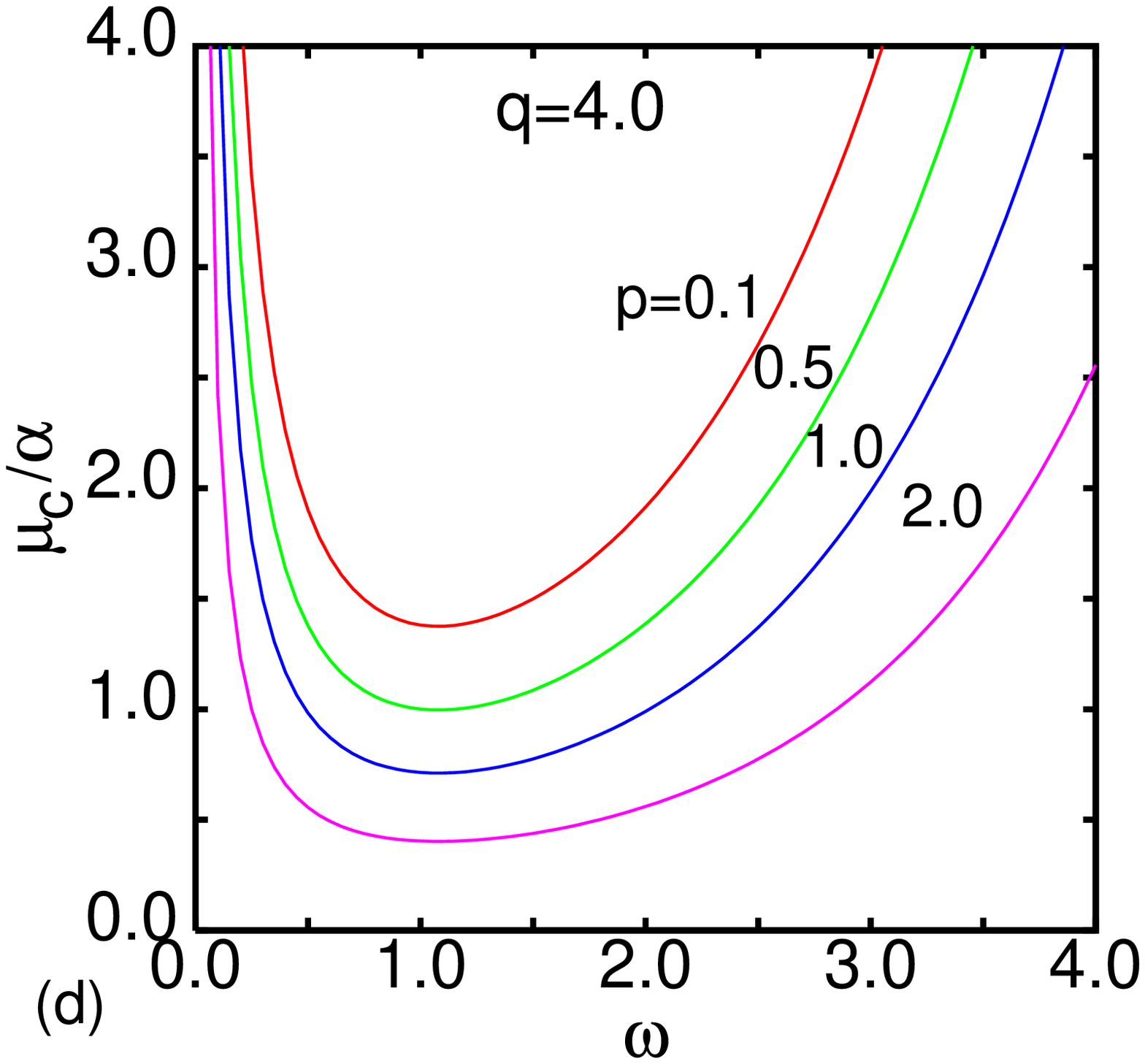,width=6.2cm,angle=-90}}

\vspace{-0.7cm}
\centerline{
\epsfig{file=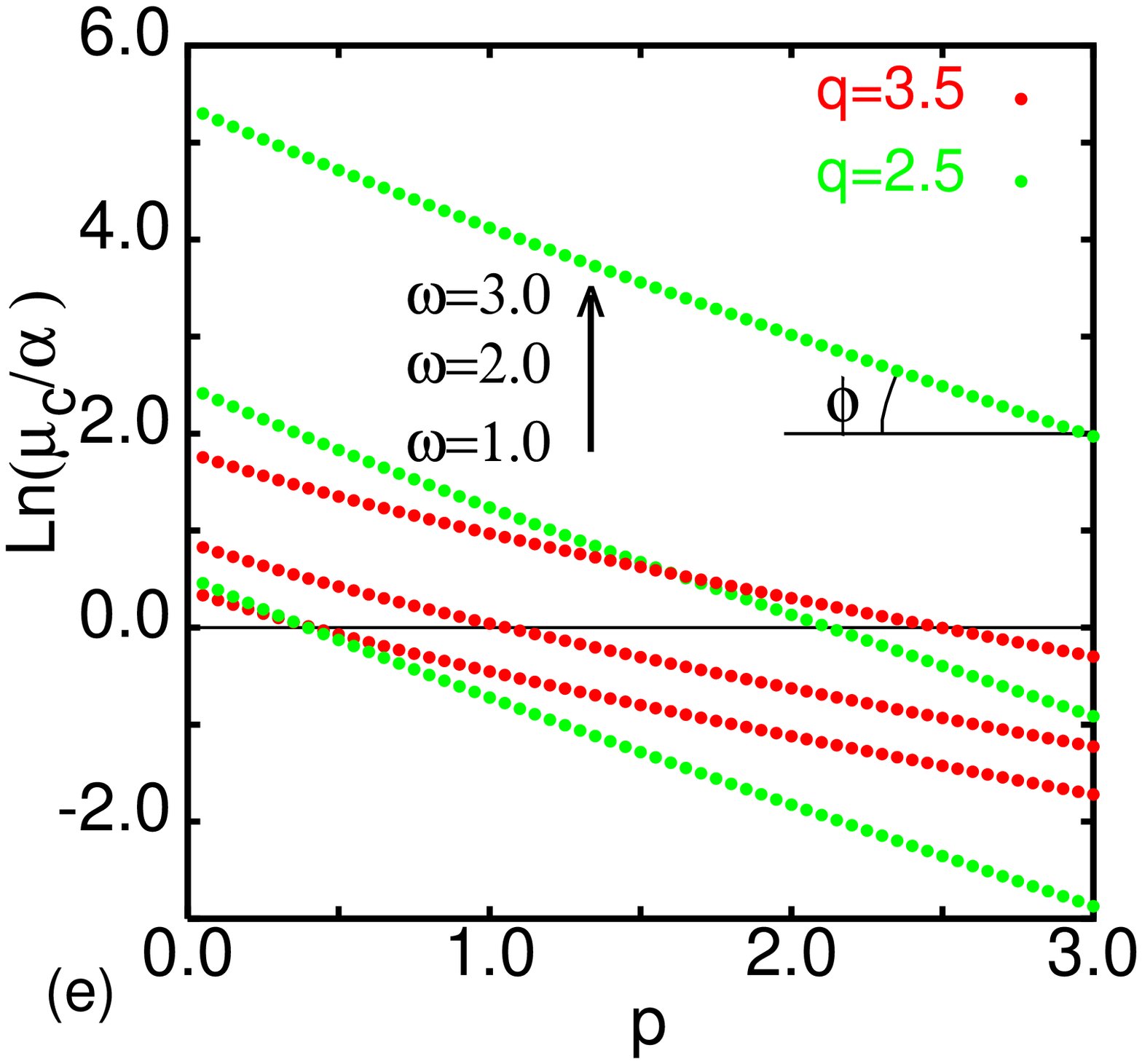,width=6.2cm,angle=-90} \hspace{0.3cm}
\epsfig{file=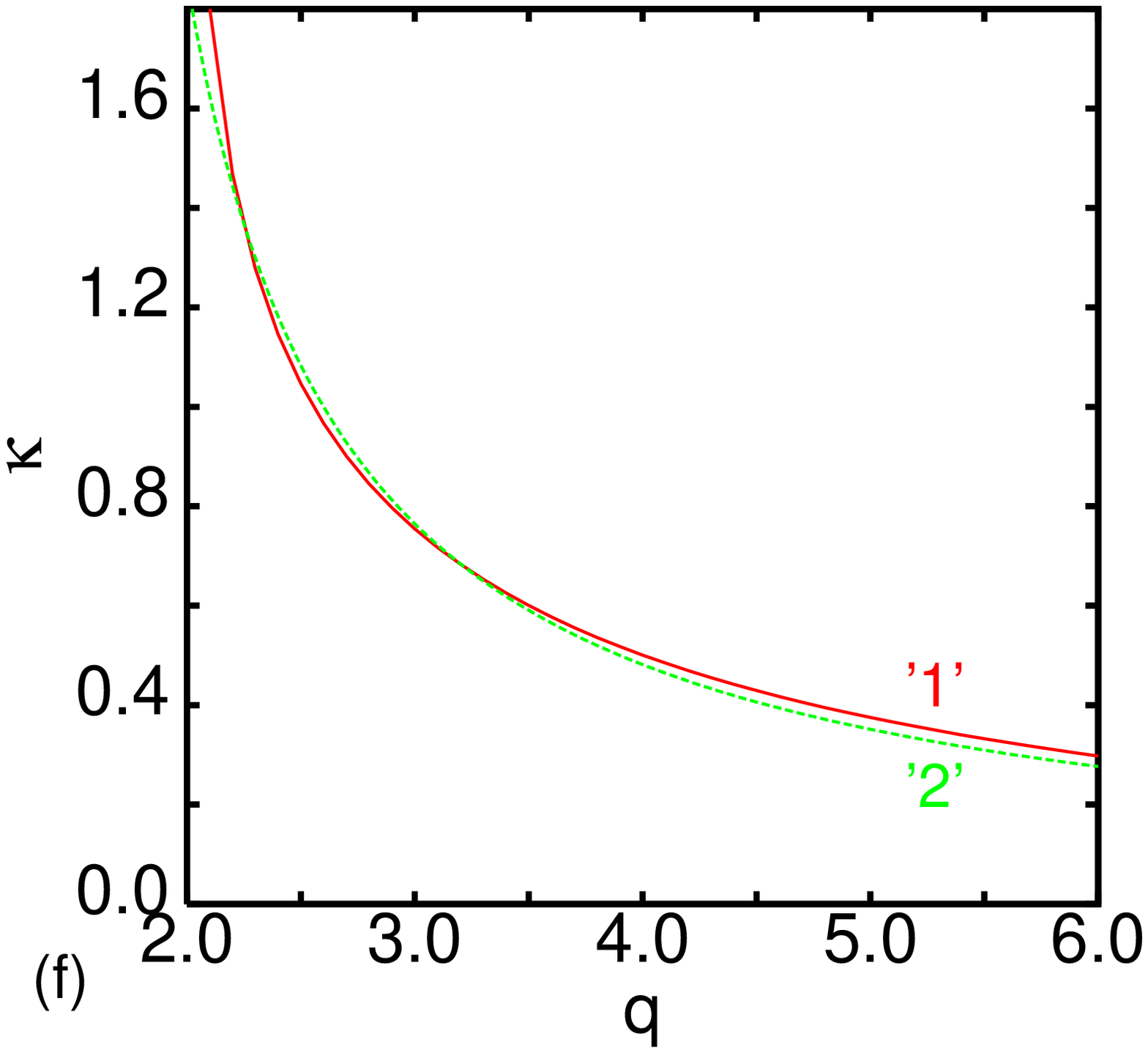,width=6.2cm,angle=-90}}

\caption{\label{fig4} Critical amplitude $\mu_c/\alpha$ versus frequency $\omega$
for
few values of $p$ ($p=0.1$ 0.5, 1.0, 2.0) and  different $q$ ($q=2.5$  in Fig. \ref{fig4}a, 
$q=3.0$ in Fig. 
\ref{fig4}b,
$q=3.5$ in Fig. \ref{fig4}c, $p=4.0$ in Fig. \ref{fig4}d.
$\ln(\mu_c/\alpha)$ versus the exponent $p$ for  three values of $\omega$ and $q=2.5$ 
and 3.5 (Fig. 
\ref{fig4}e). Dependence of the slope $\kappa$  ($\kappa=\tan \phi$  in Fig. 
\ref{fig4}e) on the exponent $q$ -- Fig. \ref{fig4}f ('1' corresponds to present 
calculations for 
$\omega=1$ while 
'2' is a fitting trail $\kappa = 1/(q/1.3 +1)$).
}
 \end{figure}

Due to the reflection symmetry of potential $V(-x)=V(x)$ (Eq. \ref{eq3})
there are two symmetric  solutions for unperturbed homoclinic orbits with 
'+' and '-' signs in Eqs. \ref{eq9}-\ref{eq10}.
A family of right hand side homoclinic orbits $(x^*,v^*)$ has been 
plotted 
in Fig. \ref{fig2}. 
In unperturbed case both stable and unstable manifolds (Poincare sections of the 
orbits are usually denoted by $W_s$ and $W_u$) 
can be identified with the orbits discussed above while perturbations would
influence them in a different way \cite{Tyrkiel2005}. Existence of cross-sections of 
between $W^S$ and $W^U$ manifolds signals
Smale's horseshoe scenario of transition to chaos.

The distance $d$ (Fig. \ref{fig3}) between them can be
estimated by 
the Melnikov function $M(t_0)$:
\begin{equation}
\label{eq11}
M(t_0) = \int_{- \infty}^{ + \infty}  h( x^*, v^*)  \wedge g( x^*,
v^*) {\rm d} t
\end{equation}
where the corresponding differential form $h$ means the gradient of unperturbed
Hamiltonian (Eq. \ref{eq3}):
\begin{equation}
\label{eq12}
h = \left(\delta x^* + \gamma  {\rm sgn}(x)|x|^{q-1}
\right) {\rm d} x  + v^* {\rm d}v,
\end{equation}
while $g$ is a perturbation  form  (Eq. \ref{eq5}) to the same Hamiltonian:
\begin{equation}
\label{eq13}
 g = \left( \tilde{ \mu} \cos{\omega t} - \tilde{\alpha} v^* \left| v^*
\right|^{p-1} \right)
{\rm d}x.
\end{equation}
All differential forms are
defined on homoclinic orbits $(x,v)=(x^*,v^*)$ (Eqs. 
\ref{eq9}-\ref{eq10}).

Thus the Melnikov function $M(t_0)$:
\begin{eqnarray}
M(t_0) &=&\int_{- \infty}^{ + \infty} v^*(t) \left( \tilde{ \mu} 
\cos{(\omega(
t+t_0))}
- \tilde{\alpha} v^{*}(t) \left| v^*(t)
\right|^{p-1} \right) {\rm d} t \nonumber \\
&=&- \sin{\omega t_0} \int_{- \infty}^{ + \infty}  v^*(t)  \tilde{ \mu}
\sin{\omega
t} {\rm d} t 
- \int_{- \infty}^{ + \infty} \tilde{\alpha} v^{* 2} (t) \left| v^*(t)
\right|^{p-1}  {\rm d} t \label{eq14} \\
&=& - \sin({\omega t_0}) \tilde \mu I_1- \tilde \alpha I_2, \nonumber
\end{eqnarray}
where $I_1$ and $I_2$ are integrals to be evaluated.
$\sin{\omega t_0}$ appears 
because of the odd parity of the function $v^*(t)$
under the above integral where
\begin{equation}
\cos(\omega (t+t_0))=\cos(\omega t)\cos(\omega t_0)-\sin(\omega 
t)\sin(\omega t_0).
\end{equation}

Thus a condition for a  global homoclinic transition, corresponding to a
hors-shoe
type of stable and unstable manifolds
cross-section (Fig. \ref{fig2}), can be written as:
\begin{equation}
\label{eq15}
{\displaystyle \bigvee_{t_0}}
~~~ M(t_0)=0 \hspace{1cm}  {\rm and}   \hspace{1cm}
\frac{\partial M(t_0)}{\partial t_0} \neq 0.
\end{equation}

For a perturbed system the above constraint together with the explicit 
form of 
Melnikov function Eq. \ref{eq14} gives the critical amplitude $\mu_c$:

\begin{equation}
\label{eq16}
\frac{\mu_c}{\alpha}= \left| \frac{I_2}{I_1} \right|,
\end{equation}
where $I_1$ and $I_2$ are corresponding integrals given in Eq. \ref{eq14}. In case 
of $I_1$ we have the following integral
\begin{equation}
\label{eq17}
I_1=\sqrt{-\delta}\left(\frac{- \delta q}{2 \gamma}
\right)^{\frac{1}{q-2}} \int_{-\infty}^{-\infty}\frac{\tanh \left(  \frac{(q-2)}{2}
\sqrt{-\delta} t
\right)}{\cosh^{\frac{2}{q-2}}
\left( \frac{(q-2)}{2}
\sqrt{-\delta} t
\right)} \sin (\omega t) \hspace{0.2cm} {\rm d} t
\end{equation}
to be evaluated numerically in general but for some cases can be easily performed numerically
(see Appendixes B and C)
while, in analogy to \cite{Trueba2000,Trueba2002,Borowiec2007}, $I_2$ can be 
expressed as  
\begin{eqnarray}
I_2 &=&
(-\delta)^{\frac{p+1}{2}}\left(\frac{- \delta q}{2 \gamma}
\right)^{\frac{p+1}{q-2}}
\int_{- \infty}^{ + \infty}
\frac{\sinh^{p+1} \left(  \frac{(q-2)}{2}
\sqrt{-\delta} t
\right)}{\cosh^{\frac{q(p+1)}{q-2}}
\left( \frac{(q-2)}{2}
\sqrt{-\delta} t
\right)}~
 {\rm d} t \nonumber \\
&=& (-\delta)^{\frac{p+1}{2}}\left(\frac{- \delta q}{2 \gamma}
\right)^{\frac{(p+1)}{q-2}} {\rm B}\left(\frac{p+2}{2}, \frac{p+1}{(q-2)}\right), 
\label{eq18} 
\end{eqnarray}
where 
B$(r,s)$ is the Euler Beta function dependent of arbitrary complex arguments with real parts
(${\rm Re}$ $r > 0$ and ${\rm Re}$ $s > 0$)
defined as
\begin{equation}
\label{eq19} 
{\rm B}(r,s)=\frac{\Gamma (r)\Gamma (s)}{\Gamma (r + s)},
\end{equation}
while $\Gamma(r)$ denotes the Euler Gamma function:
\begin{equation}
\label{eq20}
\Gamma (z) =\int_0^{\infty} {\rm e}^{-s}s^{z-1} {\rm d} s \hspace{1cm} {\rm for} \hspace{1cm} 
{\rm Re}\hspace{0.2cm}  z > 
0. 
\end{equation}

In Fig. \ref{fig4}a--d we plotted the results of Melnikov analysis for a 
critical amplitude $\mu_c/\alpha$
for
few values of $p$ ($p=0.1$ 0.5, 1.0, 2.0) and different $q$ 
($q=2.5$  in Fig. \ref{fig4}a, $q=3.0$ in Fig. \ref{fig4}b,
$q=3.5$ in Fig. \ref{fig4}c, $p=4.0$ in Fig. \ref{fig4}d)
For $\mu > \mu_c$ the system can transit to chaotic vibrations.
Note, in spite of some quantitative changes
all four figures (Fig. \ref{fig4}a-d) have similar shape and the
sequence of corresponding curves with particular exponents $p=0.1$ 0.5, 1.0, 2.0 is 
preserved
for any $\omega$ and $q$. This gives us a conviction that $p$ may play some independent role.
In fact plotting $\ln(\mu_c/\alpha)$ in Fig. \ref{fig4}e
versus $p$ and we have got  straight lines with characteristic slope independent 
on $\omega$ but changing with $q$.
Dependence of the slope $\kappa$:
\begin{equation}
\kappa=\tan \phi
\end{equation}
 defined in Fig. \ref{fig4}e, versus
$q$ has been plotted in Fig. \ref{fig4}f for $\omega=1$.
Note, the curve '1' corresponds to present calculations for
 while
'2' is a fitting curve:
\begin{equation}
\kappa = \frac{1}{\frac{q}{1.3} -1}
\end{equation}
The above scaling is not a surprise  taking into account the structure of 
$M(t_0)$ (Eq. \ref{eq12}).
In this expression the exponent $p$ is entering to the second integral independent of $\omega$.
On can also look into the analytic formulae for $\mu_c$ in 
cases of $q=4$ and $3$ in the Appendix B
(Eq. \ref{eqB.5}-\ref{eqB.8}) where the $p$ appears as an exponent.

\section{Results of Numerical Simulations}

\begin{figure}

\centerline{
\epsfig{file=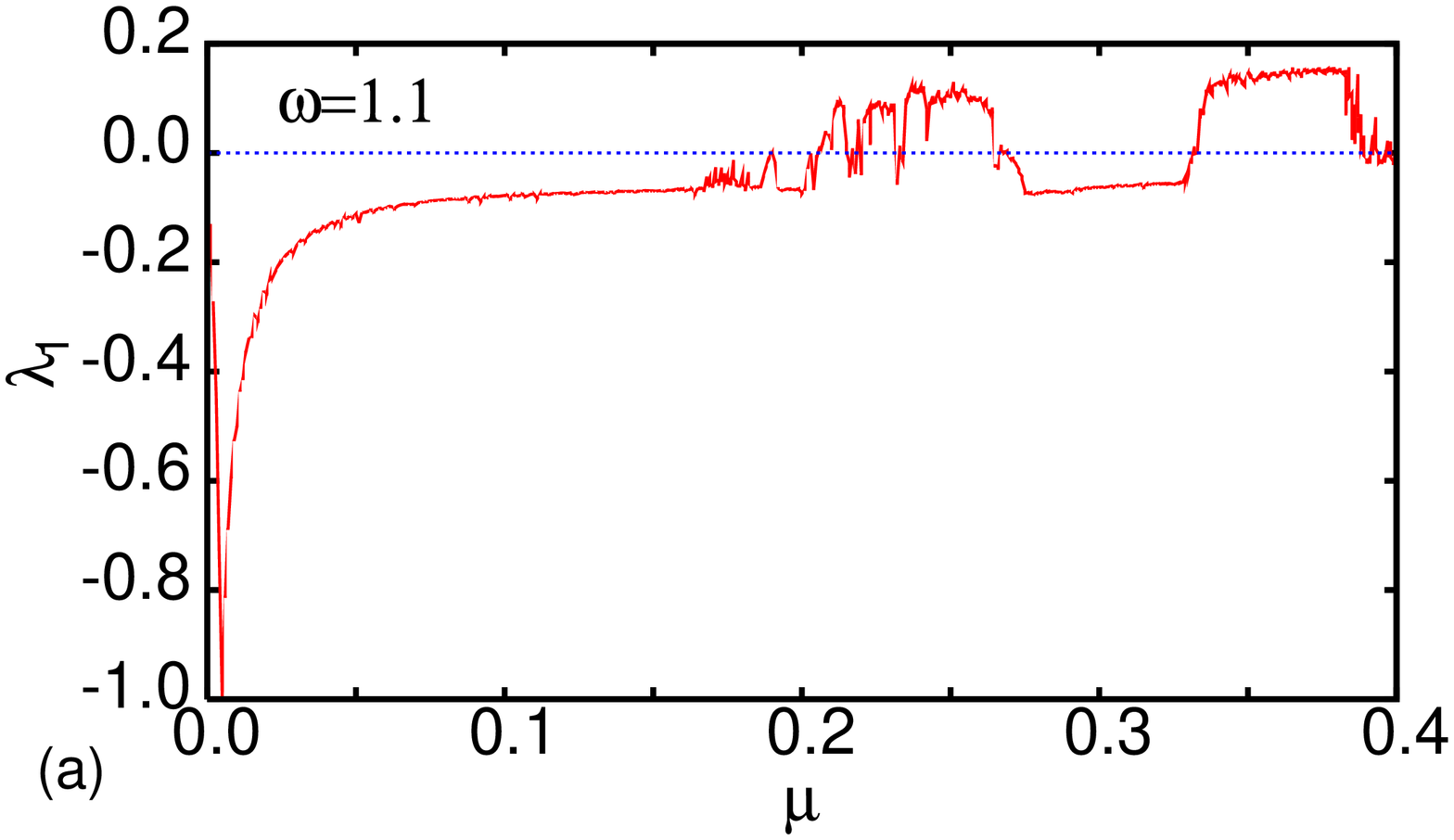,width=5.5cm,angle=-90}}

\centerline{
\epsfig{file=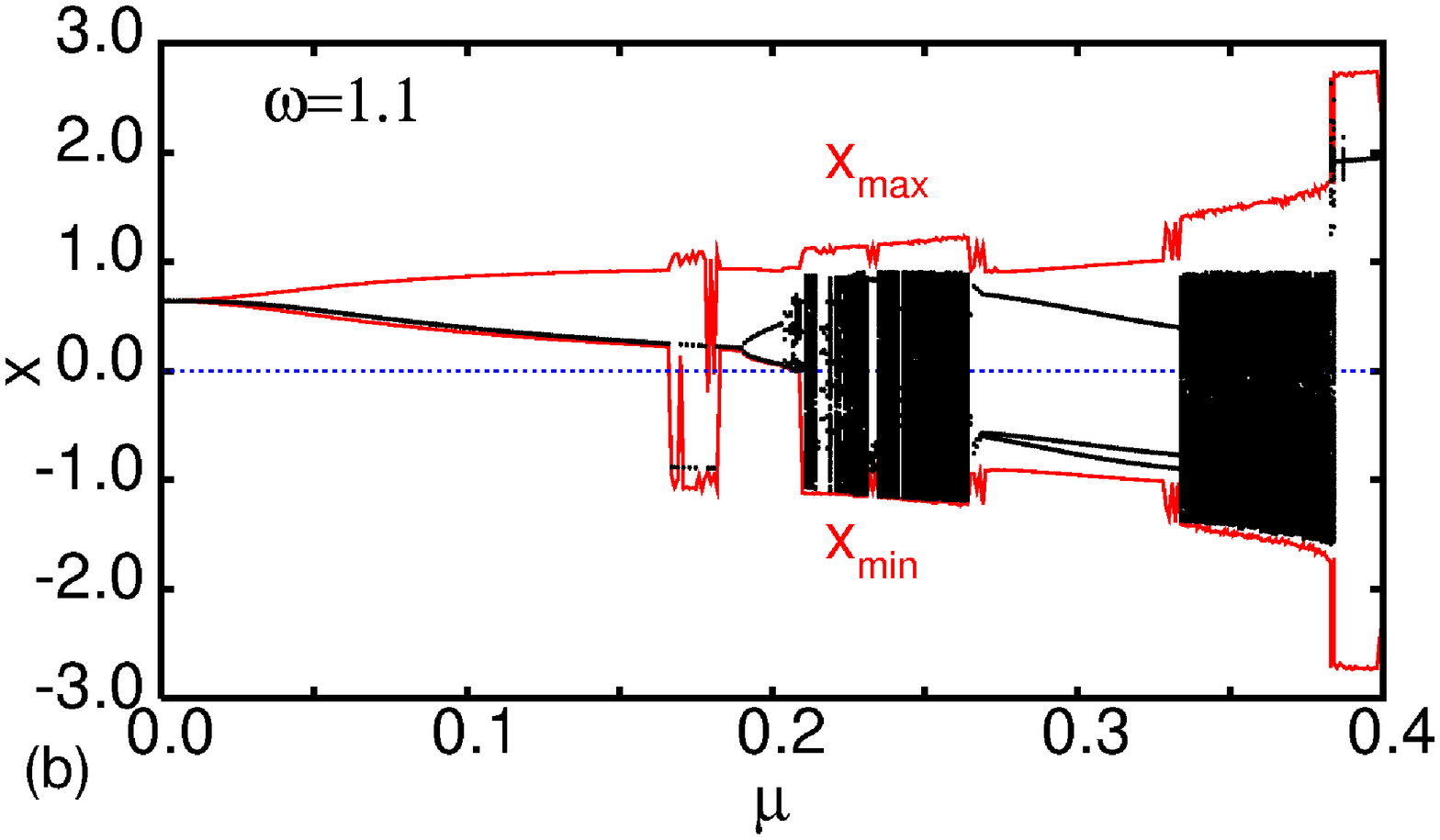,width=5.5cm,angle=-90}}

\caption{ \label{fig5} 
Maximal Lyapunov exponent $\lambda_1$ versus $\mu$  (Fig. \ref{fig5}a), 
bifurcation diagram together with size of attractor $x_{max}$ and $x_{min}$ versus  $\mu$
 (Fig. \ref{fig5}b)  for $\omega=1.1$.  }

\end{figure}

\begin{figure}

\centerline{
\epsfig{file=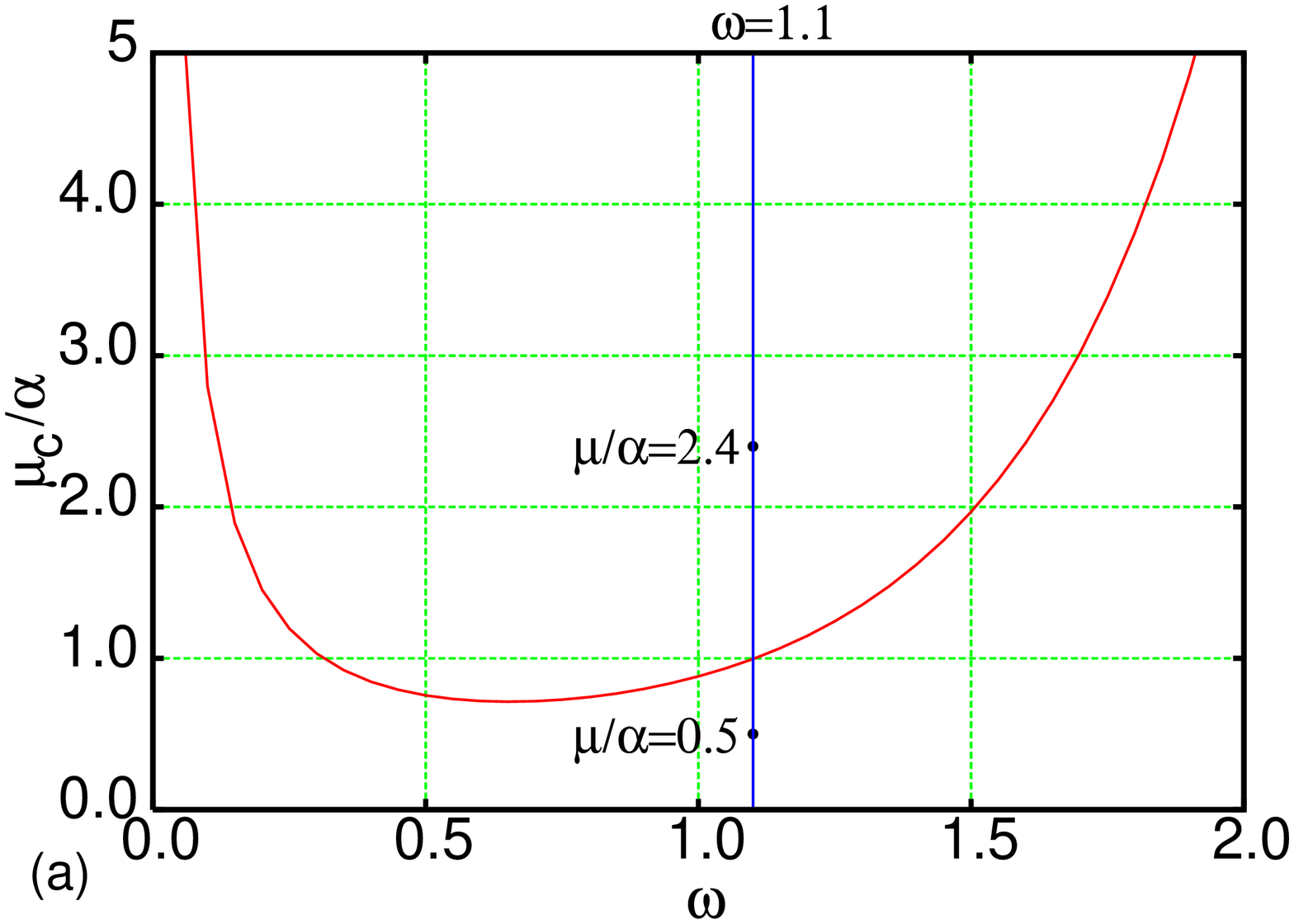,width=5.5cm,angle=-90}}

\centerline{
\epsfig{file= 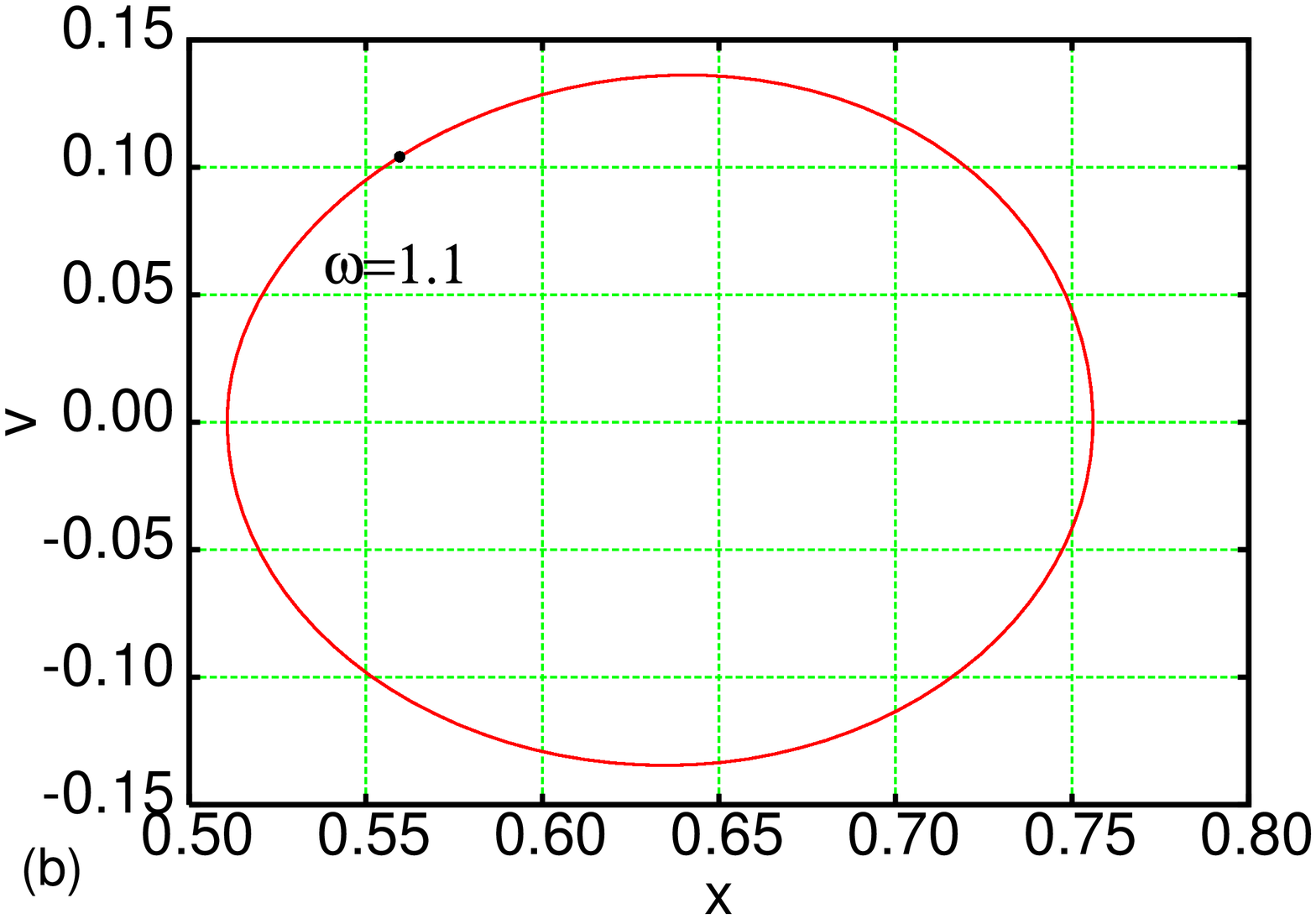,width=5.5cm,angle=-90}}

\centerline{
\epsfig{file= 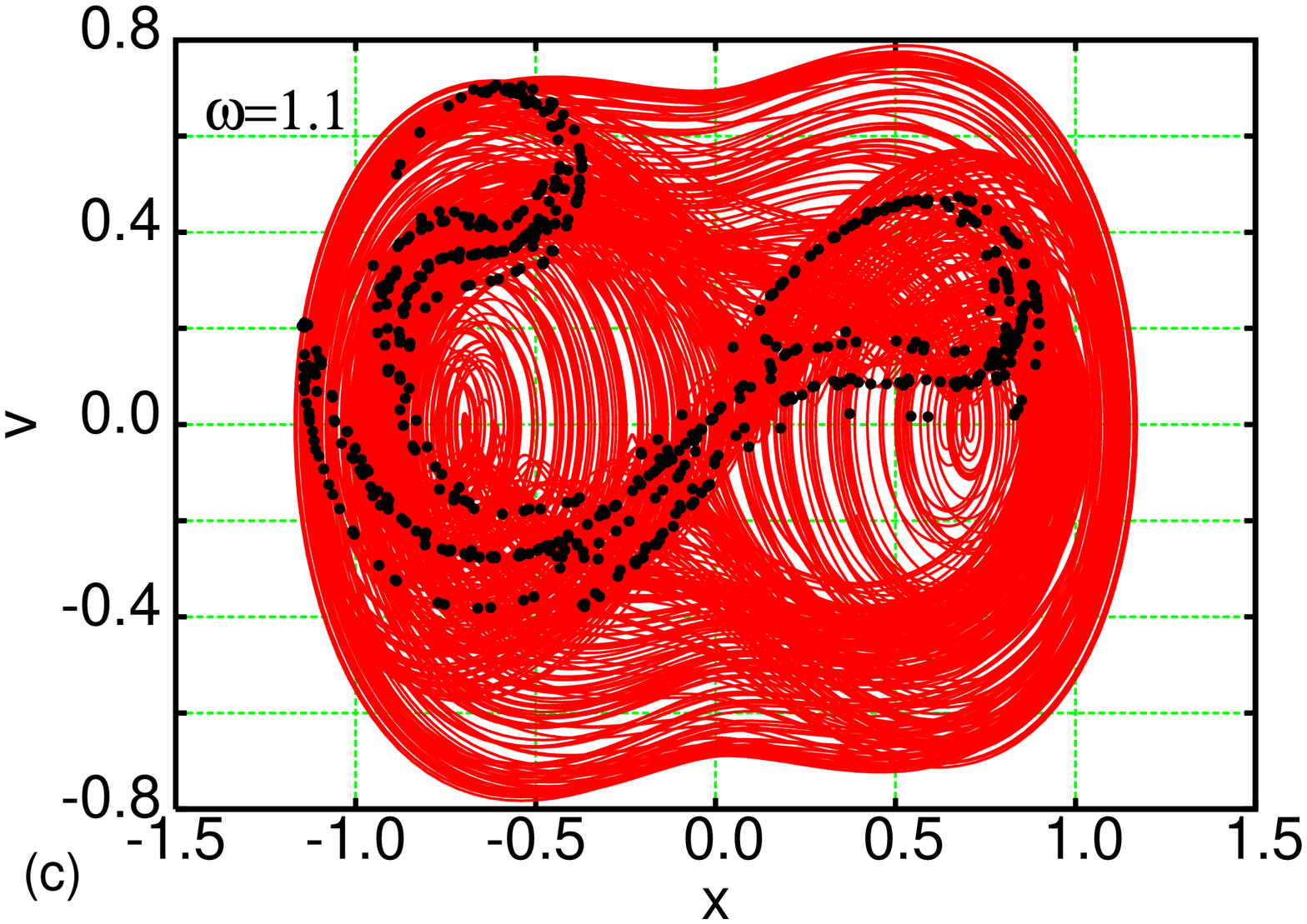,width=5.5cm,angle=-90}}

\caption{ \label{fig6} Critical amplitude $\mu_c/\alpha$ versus $\omega$ the dashed line 
corresponds to 
$\omega=1.1$ (Fig. \ref{fig6}a). Phase portrait and Poincare  maps for $\omega=1.1$, 
$\alpha=0.1$ 
and  two 
different 
$\mu$ ($\mu =0.05$ in Fig. \ref{fig6}b while  $\mu =0.24$ in Fig. \ref{fig6}c).   The 
results have been 
obtained
for $p=0.5$ and $q=2.5$.
}

\end{figure}

To illustrate the dynamical behaviour of the system it is necessary to simulate the
proper equations. Here we have used the Runge-Kutta method of the forth order and
Wolf algorithm \cite{Wolf1985} to identify the chaotic motion.
In our numerical code we started calculations from the same initial conditions 
$(x_0,v_0)=(0.45,0.1)$ for any 
new examined value of $\mu$. The system parameters $\delta=-2$ and $\gamma=q$
have be chosen the same as for  analytic calculations. We have performed 
numerical calculations for different choices of system parameters: $\alpha$, $\omega$, $p$ 
and $q$, hut here, 
for or technical 
reasons, we limited 
our discussion to  $\alpha=0.1$, $\omega=1.1$, $p=0.5$ and $q=2.5$.

In Fig. \ref{fig5}a we have plotted the maximal Lyapunov exponent versus external forcing 
amplitude $\mu$.
Here
one can clearly see points of $\lambda_1$ sing changes.
For $\mu \in [0.23, 0.27]$ and $[0.33, 0.38]$ we have got  $\lambda_1 > 0$  
 indicating  chaotic vibrations. 
In Fig. \ref{fig5}b we have plotted 
the corresponding bifurcation diagram.
Thus a black region means  chaotic motion.   
This result, as well as others, calculated for different sets of system parameters, is 
consistent with the Melnikov results. For comparison we have plotted 
the Melnikov curve again (Fig. \ref{fig6}a with two trial points 
$\mu=0.05$ and $\mu=0.24$
(for $\alpha=0.1$).  There is no doubt that Fig. \ref{fig6}b 
shows the regular synchronized motion 
represented by a single loop on a phase portrait and a singular point on
Poincare stroboscopic map.  
On the other hand
Fig. \ref{fig6}c shows clearly a strange attractor of chaotic vibration with complex 
structure of the Poincare map.

\section{Summary and Conclusions} 
We have examined criteria for transition to
chaotic
vibrations
in the double well system with a  damping term ${\rm dpt} (v)=v|v|^{p-1}$  described by a  
fractional 
exponent $p$ and nonlinear potential with negative square term 
(related negative stiffness) and a positive term with 
higher
exponent $|x|^q$ where $q > 2$.
In spite of non-smoothness of corresponding vector-fields ($h$ and $g$ -- Eqs. \ref{eq11} and
\ref{eq12}, respectively) it has been proven in the Appendix A that extra terms to the 
Melnikov integral \cite{Kunze2001} projected out.
Thus the critical value of 
excitation amplitude $\mu$ above which the system vibrate chaotically has been estimated, in 
 by 
means of the Melnikov theory \cite{Melnikov1963}.
For some selected values of the exponent $q$ ($q=$4,3, 8/3, 2/5) it was possible to
derive a final formula for $\mu_c$  while for 
other cases one of  Mielnikov' integrals has be calculated numerically. 

The analytical results have been confirmed by simulations. In this approach 
we used standard methods of analysis as Poincare maps, bifurcation diagrams and 
Lyapunov exponent. 

The Melnikov method, is sensitive to a
global homoclinic bifurcation 
and gives a necessary condition for excitation amplitude $\mu=\mu_{c1}$ system 
in its transition to 
 chaos \cite{Guckenheimer1983,Wiggins1990}. On the other hand the largest Lyapunov exponent 
\cite{Wolf1985}, measuring the local 
exponential 
divergences 
of particular phase portrait trajectories gives a 
sufficient  condition $\mu=\mu_{c2}$  for this 
transition
which has obviously a higher value of the excitation amplitude $\mu=\mu_{c2} > 
\mu_{c1}$.

Above the Melnikov transition predictions ($\mu > \mu_{c1}$) 
we have obtained transient chaotic vibrations 
\cite{Wiggins1990,Szemplinska1993,Szemplinska1995,Tyrkiel2005,Borowiec2007}
as we expected
drifting to a regular steady state away the fractal attraction regions separation  boundary. 
This is typical behaviour of the system which undergo
global homoclinic bifurcation.

\section*{Acknowledgments}
This paper has been partially  supported by the  Polish Ministry of 
Science and Information. 
GL would like to thank  Max Planck 
Institute for the Physics of Complex Systems in Dresden for hospitality.

\vspace{1cm}

\appendix{\large  \hspace{1.5cm} \bf \noindent Appendix A} \\~\\
\def\thesection{A}
\setcounter{equation}{0}
\def\theequation{A.\arabic{equation}}  

\setcounter{figure}{0}
\def\thefigure{A.\arabic{figure}}  %

Starting with the perturbation equation (Eq. \ref{eq6}) we  write it in a two element 
vector form

\begin{equation}
\label{eqA.1}
\dot {\bf q} = {\bf h}+ \epsilon {\bf g},
\end{equation}
where
\begin{eqnarray}
{\bf q} &=&[x,v] \nonumber \\
{\bf h} &=& [v,- \delta x -\gamma {\rm sgn}(x) |x|^{q-1}]  \label{eqA.2}\\
{\bf g} &=& [0,-\tilde{\alpha} v \left| v \right|^{p-1}+\tilde{\mu} \cos{ \omega t}]. 
\nonumber 
\end{eqnarray}
On the other hand the homoclinic orbit
\begin{equation}
\label{eqA.3}
{\bf q}^*(t-t_0)= [x^*(t-t_0),v^*(t-t_0)],
\end{equation}
where $t_0$ is usually defined by simple zero of Melnikov integral \cite{Melnikov1963}.  
In the limit of extreme time 
$t \rightarrow  \pm \infty$  the system state $[x,v]$ reaches a saddle point
 $[x,v]=[0,0]$  
(see Figs. \ref{fig1}a, 
\ref{fig2}).
Consequently, in the aim to examine the Melnikov criterion for chaos appearance, the vector 
fields 
${\bf h}$ and ${\bf g}$ are defined on the homoclinic orbit (Fig. \ref{fig2}) as:
\begin{equation}
\label{eqA.4}
{\bf h}(q^*)= [v^*(t-t_0),- \delta x^*(t-t_0) -\gamma {\rm sgn}(x^*(t-t_0) 
|x^*(t-t_0)|^{q-1}]
\end{equation}
and
\begin{equation}
\label{eqA.5}
{\bf g}(q^*,t) = [0,-\tilde{\alpha} v^*(t-t_0) \left| v^*(t-t_0) \right|^{p-1}+\tilde{\mu} 
\cos{ 
\omega t}].
\end{equation}
The perturbed stable and unstable manifolds $W^s$ and $W^u$ read 
\cite{Kunze2001}
\begin{equation}
\label{eqA.6}
{\bf q}^{u,s}(t,t_0)={\bf q}^*(t-t_0)+ \epsilon {\bf q}^{u,s}_1(t,t_0) + 
{\bf O}(\epsilon^2)
\end{equation}
respectively.

Note the perturbation correction to the homoclinic orbit ${\bf q}^{u,s}_1(t,t_0)$
in the above expression (Eq. \ref{eqA.6}) should be found  by solving the following linear 
differential equation
about the examined time $t$ (or a system state ${\bf q}={\bf q}^*(t-t_0)$):
\begin{equation}
\label{eqA.7}
\dot {\bf q}^{u,s}_1(t,t_0)= \left(\frac{\partial h_1}{\partial v} 
-\frac{\partial h_2}{\partial x} \right)_{|{\bf q}={\bf q}^*(t-t_0)} {\bf q}^{u,s}_1(t,t_0)+ 
{\bf 
g}({\bf 
q}^*(t-t_0),t)  
\end{equation}

\begin{figure}[htb]   
\centerline{
\epsfig{file=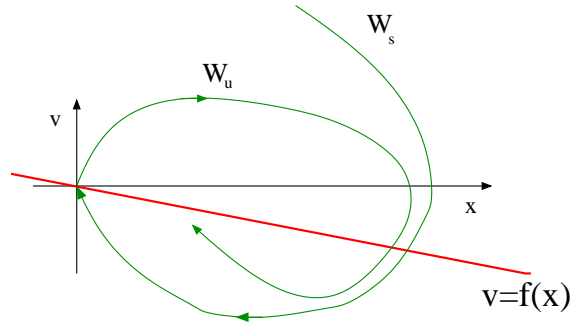,width=7.5cm,angle=0}}
\caption{ \label{figA.1}
Stable and unstable manifolds and one of lines of discontinuity $v=f(x)$. This line crosses
manifold at $[x_d,v_d]$ for the specific 
time $t=t_d$ $x_d=x(t_d)$ and $v_d=v(t_d)$. 
}
\end{figure}

 Note that the above vector fields ${\bf h}(x,v)$ and ${\bf g}(x,v,t)$ are 
not $C^2$ functions.
 Namely {\bf h} is  of $C^2$ only if  $q \ge 2$ and of $C^1$ if $1 \le q < 
2$. 
 Similarly {\bf g} is of $C^2$ only if $p=1$ or $p \ge 2$ and of $C$ if $0 
< p < 1$ and  $C^1$   
$1 < p < 2$, respectively. 
In case of $1 \le q < 2$
the line 
$x=0$ separates the whole phase space $(x,v)$ 
into two parts where ${\bf h}(x,v)$ is enough smooth ($C^2$).
The same can be applied to the line $v=0$ and  ${\bf g}(x,v,t)$
as a possible set for $0 < p< 1$ and $1 < p < 2$.
This line crosses     
manifold at $[x_d,v_d]$ for the specific
time $t=t_d$ such that $x_d=x(t_d)$ and $v_d=v(t_d)$ (Fig. \ref{figA.1}).

According to Kunze and K\"{u}pper \cite{Kunze2001} the ($C^2$) space 
separation includes additional terms to the Melnikov integral.

Thus
the Mielnikov function M($t_0$):

\begin{eqnarray}
{\rm M}(t_0)={\rm M}_0(t_0) &+& \sum_{t_d} 
\left(
{\bf h}^{\perp,+}({\bf q}^*(t_d^-)) \cdot  {\bf
q}_1^{u,+} (t_0+t_d^-,t_0) - 
{\bf h}^{\perp,-}({\bf q}^*(t_d^-)) \cdot  {\bf
q}_1^{u,-} (t_0+t_d^-,t_0) 
\right. \label{eqA.8}
\\
&+& \left. 
{\bf h}^{\perp,-}({\bf q}^*(t_d^+)) \cdot  {\bf
q}_1^{s,-} (t_0+t_d^+,t_0) -
{\bf h}^{\perp,-}({\bf q}^*(t_d^+)) \cdot  {\bf
q}_1^{s,+} (t_0+t_d^+,t_0)
\right), \nonumber 
\end{eqnarray}
where
${\bf q}_1^{s,\pm}$, ${\bf q}_1^{u,\pm}$ are stable and unstable manifold 
perturbation solutions (Eq. \ref{eqA.7}) for $t$ in the vicinity of $t_d$ but
$t > t_d$ for '$+$' sign and $t < t_d$ for '$-$' sign, respectively.

${\rm M}_0(t_0)$ is defined as for smooth vector fields: 
\begin{equation}
\label{eqA.9}
{\rm M}_0(t_0)=\int_{-\infty}^{\infty}  {\bf h}^{\perp} ({\bf q}^*(t-t_0)\cdot {\bf
g}({\bf q}^*(t+t_0),t)~{\rm d} t
\end{equation}
and ${\bf h}^{\perp}=[-h_2,h_1]$. 

Once  first discontinuity are identified for $x=0$, $t_d \rightarrow \pm \infty$,
one have to examine $\phi_1^{\pm}(t,t_0)=q_1^{u,s,\pm}(t,t_0)$
and  $\phi_2^{\pm}(t,t_0)=q_1^{u,s,\pm}(t,t_0)$: 

\begin{equation}
\label{eqA.10} 
\left\{ \begin{array}c
\dot \phi_1=(1+\delta + \gamma (q-1) x^{q-2}) \phi_1 \\
\dot \phi_2=(1+\delta + \gamma (q-1) x^{q-2}) \phi_2 - \tilde{\alpha} v |v|^{p-1} 
+\tilde{\mu} 
\cos \omega t
\end{array} \right.
\hspace{1cm} {\rm for } \hspace{1cm} x> 0 
\end{equation}

and 
\begin{equation}
\label{eqA.11}
\left\{ \begin{array}c
\dot \phi_1=(1+\delta + \gamma (q-1) (-x)^{q-2}) \phi_1 \\
\dot \phi_2=(1+\delta + \gamma (q-1) (-x)^{q-2}) \phi_2 - \tilde{\alpha} v |v|^{p-1} 
+\tilde{\mu}
\cos \omega t
\end{array} \right.
\hspace{1cm} {\rm for } \hspace{1cm} x< 0
\end{equation}
Note substituting $x=0$ for $q > 2$ to Eqs. \ref{eqA.10} and \ref{eqA.11}  we get the same 
equations for $\phi_{1/2}$ and consequently the same expressions. This means 
 automatically
no 
extra terms to the Melnikov integral (Eq. \ref{eqA.8}) caused by the $x=0$ discontinuity. 
Interestingly 
$q \le 2$ would lead to a different result but for such case the is no homoclinic orbit
in the unperturbed system described by $H^0$ (Eqs. \ref{eq5},\ref{eq3}).  

Let us non focus on $v=0$ discontinuity.
In this case
\begin{equation}
\label{eqA.12}
\left\{ \begin{array}c
\dot \phi_1=(1+\delta + \gamma (q-1) |x^*|^{q-2}) \phi_1 \\
\dot \phi_2=(1+\delta + \gamma (q-1) |x^*|^{q-2}) \phi_2 - \tilde{\alpha} (v^*)^p
+\tilde{\mu}
\cos \omega t
\end{array} \right.
\hspace{1cm} {\rm for } \hspace{1cm} v > 0
\end{equation}

\begin{equation}
\label{eqA.13}
\left\{ \begin{array}c
\dot \phi_1=(1+\delta + \gamma (q-1) |x^*|^{q-2}) \phi_1 \\
\dot \phi_2=(1+\delta + \gamma (q-1) |x^*|^{q-2}) \phi_2 + \tilde{\alpha} (v^*)^p        
+\tilde{\mu}
\cos \omega t
\end{array} \right.
\hspace{1cm} {\rm for } \hspace{1cm} v < 0
\end{equation}

Note, excluding natural odd numbers for the $p$  exponent, the above equations (Eqs. 
\ref{eqA.12} and \ref{eqA.13}) are usually different for any other $p \ge 0$.
However both solutions $[\phi_1^-,\phi_2^-]$ and $[\phi_1^+,\phi_2^+]$
have to be projected into 
\begin{equation}
\label{eqA.14}
{\bf h}^{\perp}_{|v=0}=[-h_2,h_1]_{|v=0}= [ \delta x^*(t_d-t_0) +\gamma {\rm 
sgn}(x^*(t-t_0)
|x^*(t_d-t_0)|^{q-1},0]
\end{equation}
and the differences in solutions in $\phi_2^-$ and $\phi_2^+$ are effectively 
projected out. Interestingly this is also valid for $p=0$ (a dry friction case). 

Finally for $q > 2$ and $p \ge 0$ the Mielnikov function M($t_0$) can be treated as a
\begin{equation}
\label{eqA.15}
{\rm M}(t_0)={\rm M}_0(t_0).
\end{equation}
\vspace{1cm}
~

\appendix{\large  \hspace{1.5cm} \bf \noindent Appendix B} \\~\\
\def\thesection{B}
\setcounter{equation}{0} 
\def\theequation{B.\arabic{equation}}  

In this appendix we show how to get homoclinic orbits and analytically for 
some specific cases of  exponent $q$: $q=4$, 
3, 2.67 and 2.5.

In case of  $q=4$ we follow works by Trueba {\em et al.}  
\cite{Trueba2000} and Borowiec {\em et al.} \cite{Borowiec2007} (and Eqs. 
\ref{eq9}-\ref{eq10}) 
\begin{eqnarray}
x^* =x^*(t-t_0)&=& \pm \sqrt{\frac{-2 \delta}{\gamma}}~~
\frac{1}{\cosh \left( \sqrt{-\delta} (t-t_0) \right)}
\nonumber \\
v^* =v^*(t-t_0)&=& \pm
\sqrt{\frac{2}{\gamma}}~\delta~~\frac{\tanh \left( \sqrt{-\delta} (t-t_0) 
\right)}{\cosh \left(
\sqrt{-\delta} (t-t_0) \right)}
\label{eqB.1}
\end{eqnarray}
where '$+$' and '$-$' signs are related to left-- and right--hand side 
orbits, 
respectively, $t_0$ is a time like integration constant.

On the other hand for $q=3$
we have
\begin{eqnarray}
x^*=x^*(t-t_0) &=& \mp \frac{3 \delta}{2 \gamma}~~ \frac{1}
{\cosh^2 \left( \frac{\sqrt{-\delta} (t-t_0)}{2}\right)}
\nonumber \\
v^*=v^*(t-t_0) &=& \mp \frac{3 \delta \sqrt{-\delta}}{2 \gamma}
~~ \frac{\tanh \left(
\frac{\sqrt{-\delta} (t-t_0)}{2}\right)}{ \cosh^2 \left(
\frac{\sqrt{-\delta} (t-t_0)}{2}\right)},  
\label{eqB.2}
\end{eqnarray}

Consequently for $q=2\frac{2}{3}=8/3\approx 2.67$
we have
\begin{eqnarray}
x^*=x^*(t-t_0) &=& \pm \left( \frac{-4 \delta}{3 \gamma} \right)^{3/2}~~
\frac{1}
{\cosh^3 \left( \frac{\sqrt{-\delta} (t-t_0)}{3}\right)}
\nonumber \\
v^*=v^*(t-t_0) &=& \mp \left( \frac{-4 \delta}{3 \gamma}\right)^{3/2} 
~\sqrt{-\delta}~~ \frac{\tanh \left(
\frac{\sqrt{-\delta} (t-t_0)}{3}\right)}{ \cosh^3 \left(
\frac{\sqrt{-\delta} (t-t_0)}{3}\right)},
\label{eqB.3}
\end{eqnarray}

And for $q=2.5$
\begin{eqnarray}
x^*=x^*(t-t_0) &=& \mp \left( \frac{5 \delta}{4 \gamma} \right)^2~~
\frac{1}
{\cosh^4 \left( \frac{\sqrt{-\delta} (t-t_0)}{4}\right)}
\nonumber \\
v^*=v^*(t-t_0) &=& \pm \left( \frac{4 \delta}{3 \gamma}\right)^2 
~\sqrt{-\delta}~~ \frac{\tanh \left(
\frac{\sqrt{-\delta} (t-t_0)}{4}\right)}{ \cosh^4 \left(
\frac{\sqrt{-\delta} (t-t_0)}{4}\right)},
\label{eqB.4}
\end{eqnarray}

The results for a Melnikov integral can be easily found in the above cases.
Evaluating the corresponding integral (Eq. \ref{eq11})
after some algebra the last condition (Eq. \ref{eq15}) yields to
a critical value of excitation amplitude $\mu_c$.
Thus for  $q=4$ \cite{Moon1979,Trueba2000,Trueba2002,Borowiec2007} we have:
\begin{equation}
\label{eqB.5}
\mu_c= \alpha \frac{2^{p/2}
(-\delta)^{p+1/2}}{\pi \omega \gamma^{p/2}}
{\rm B} \left(
\frac{p+2}{2},\frac{p+1}{2}\right)
\cosh \left( \frac{\pi \omega}{2 \sqrt{-\delta}}\right), 
\end{equation}
while in case of  $q=3$ \cite{Thompson1989,Litak2005,Litak2007}:
\begin{equation}
\label{eqB.6}
\mu_c=\alpha \frac{3^p (-\delta)^{3p/2+2}}{2^{p+1} \pi \omega^2 \gamma^p} 
{\rm B} \left(                                 
\frac{p+2}{2},p+1\right)
\sinh \left( 
\frac{\pi \omega}{\sqrt{-\delta}},
\right) 
\end{equation}
for $q=8/3$:
\begin{equation}
\label{eqB.7}
\mu_c= \alpha \frac{2^{\frac{6}{5}(p+1)} (-\delta)^{11p/10-2/5}}
{3^{\frac{3}{5}(p+1)}\pi(9 \omega^2-\delta) \omega \gamma^{3p/5+9/10}} {\rm B} \left(
\frac{p+2}{2},\frac{3(p+1)}{2}\right) \cosh
\left(\frac{3\pi\omega}{2\sqrt{-\delta}}\right),
\end{equation}
and finally for $q=5/2$: 
\begin{equation}
\label{eqB.8}
\mu_c= \alpha \frac{5^{2p} (-\delta)^{5p/2+3/2}}{2^{4p+3} \pi (4 \omega^2 -\delta) 
\omega^2\gamma^{2p}}{\rm B} \left(
\frac{p+2}{2},2(p+1)\right) \sinh \left(\frac{2\pi\omega}{\sqrt{-\delta}} \right).
\end{equation}

\vspace{1cm}

\appendix{\large  \hspace{1.5cm} \bf \noindent Appendix C} \\~\\
\def\thesection{C}
\setcounter{equation}{0}
\def\theequation{C.\arabic{equation}}  

\setcounter{figure}{0}
\def\thefigure{C.\arabic{figure}}  

The integral $I_1$ can be evaluated analytically in some specific cases of
exponents $q$ corresponding to homoclinic orbits Eqs. \ref{eqB.1}-\ref{eqB.4}
numbered by the corresponding power index $m$ applied to hyperbolic $\cos$ function in the 
denominators.
Let us consider integrals $I_1$
 for given $m=1$,2,3 and 4 related to various $q$ exponents $q=4$, 3, 8/3 
and 5/2, respectively.    To better clarity we will use new notation $I_1 \rightarrow I_1(m)$ for 
given $m$:
\begin{eqnarray}
\label{eqC.1}
I_1(m)=\int_{- \infty}^{ + \infty}  v^*(t)  \tilde{ \mu}
\sin{\omega
t} {\rm d} t =C_m \int_{- \infty}^{ + \infty} \frac{\tanh(\tau)}{\cosh^{m}(\tau)} 
\sin(\omega_m \tau)  {\rm d} \tau \\
=\frac{\omega_mC_m}{m} \int_{- \infty}^{ + \infty} \frac{\cos(\omega_m 
\tau)}{\cosh^m(\tau)}  
{\rm d} \tau= 
\frac{\omega_mC_m}{m} J_m (\omega_m), \nonumber 
\end{eqnarray}
while constants $C_m$ and $\omega_m$ are defined as follows:
\begin{equation}
\label{eqC.2}
C_m=\sqrt{-\delta} \left(\frac{-(m+1)\delta}{m\gamma} \right)^{m/2} , \hspace{1.2cm} 
\omega_m=
\frac{m\omega}{\sqrt{-\delta}}
\end{equation}
Evaluating the integral  $J_m(\omega_m)$ (\ref{eqC.1}), for positive integer $m$, twice 
by parts
we have got the following recurrence identity
\begin{equation}
\label{eqC.3}
J_{m+2}(\omega_{m+2})= \frac{\omega_{m+2}^2 +m^2}{m(m+1)} J_m (\omega_{m+2}) \hspace{1cm} {\rm 
for} \hspace{1cm} 
m=1,2,3,...
\end{equation}

Thus only $J_1$ and $J_2$ need to be calculated. Below we evaluate them 
 on the complex plane by summing corresponding 
residue.

\begin{figure}[htb]
\centerline{
\epsfig{file=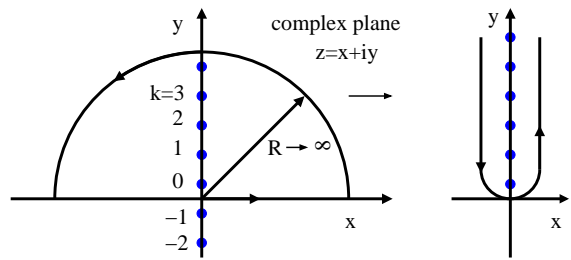,width=7.5cm,angle=0}}
\caption{ \label{figC.1}
Deformed contour integration schema and imaginary poles.
}
\end{figure}

\begin{equation}
\label{eqC.4}
\oint f(z) {\rm d} z = 2 \pi {\rm i} \sum_{k=1}^N
{\rm Res}[f(z),z_k],
\end{equation}
where
\begin{equation}
\label{eqC.5}
{\rm Res}[f(z),z_k]= \frac{1}{(m-1)!} \lim_{z \rightarrow z_k}
\frac{{\rm d}^{ m-1}}{{\rm d} z^{m-1}} \left[(z-z_k)^mf(z)\right],
\end{equation}
for $m=$ 1 or 2, in our case.

The examined function $f(z)$ is defined as:
\begin{equation}
\label{eqC.6}
f(z)=\frac{2^m}{(\exp (z) + \exp( -z))^m} \exp {({\rm i} \omega_m z )}.
\end{equation}
Note that  
 on the real axis (Fig. \ref{figC.1}) ${\rm Re}$ $z=\tau$ it can be written as
\begin{equation}
\label{eqC.7}
{\rm Im} f(\tau)= \frac{\cos (  \omega_m \tau)}{\cosh^m
\tau}.
\end{equation}
The multiplicity  of each pole of the complex function $f(z)$ (Eq. \ref{eqC.6}) is 
given by
\begin{equation}
\label{eqC.8}
z_k=\left( \frac{\pi}{2} + \pi k \right) {\rm i} \hspace{1cm} {\rm for} \hspace{1cm} 
k=1,2,3,...
\end{equation}
Note $J_m$ (Eq. \ref{eqC.1}) can be easily determined for $m=1$ or 2. Namely, after 
summation of all 
poles in the upper half-plane 
(Fig. 
\ref{figC.1}), we get for $m=1$
\begin{equation}
\label{eqC.9}
 J_1=\int_{-\infty}^{+\infty} {\rm d} \tau  \frac{ \cos( \omega_1 
\tau)}{\cosh \tau} =
\frac{\pi}{\cosh \left(\frac{\pi
\omega_1}{2}
\right)} 
\end{equation}
while for  $m=2$ we obtain
\begin{equation}
\label{eqC.10}
J_2=\int_{-\infty}^{+\infty} {\rm d} \tau  \frac{
\cos( \omega_2 \tau)}{\cosh^2 \tau} =
\frac{\pi \omega_2}{\sinh \left(\frac{\pi
\omega_2}{2}
\right)}.
\end{equation}

On the other hand, in case of  $m=3$ and $m=4$  (and also for any larger $m$), we can use the 
recurrence relation 
(Eq. \ref{eqC.3}):
\begin{equation}
\label{eqC.11}
J_3=
\frac{\pi (\omega_3^2+1) }{2\cosh \left(\frac{\pi\omega_3}{2}\right)},
\hspace{2cm}
J_4=\frac{\pi \omega_4(\omega_4^2+4 ) }{6\sinh \left(\frac{\pi
\omega_4}{2} \right)}.
\end{equation}
Consequently using Eq. \ref{eqC.1}

\begin{eqnarray}
&& I_1(1)= \left( \frac{-2 \delta}{\gamma} \right)^{1/2}
\frac{\pi \omega}{ \cosh \left(\frac{\pi\omega}{2\sqrt{-\delta}}\right)},
\hspace{2.3cm} 
I_1(2)=  \left( \frac{-3 \delta}{2\gamma} \right)
 \frac{2\pi \omega^2}{\sqrt{-\delta} \sinh
\left(\frac{\pi
\omega}{\sqrt{-\delta}}
\right)}, 
\\
&& I_1(3)=  
\left(\frac{-4 \delta}{3 \gamma}\right)^{3/2} 
\frac{\pi (9\omega^2-\delta) }{2 
\sqrt{-\delta} \omega \cosh
\left(\frac{3\pi\omega}{2\sqrt{-\delta}}\right)},
\hspace{1.5cm}
I_1(4)=\left(\frac{-5 \delta}{4 \gamma} \right)^2
\frac{8\pi (4\omega^2-\delta )}{3(-\delta)\sinh 
\left(\frac{2\pi\omega}{\sqrt{-\delta}} \right)}
. \nonumber
\end{eqnarray}

\end{document}